\documentstyle[referee]{l-aa}

\newcommand{\gray}{{$\gamma$-ray}}
\newcommand{\grays}{{$\gamma$-rays}}
\newcommand{\etal}{{\it et al.}}

\newcommand{\xxiicrc}{{\it 21st Internat. Cosmic Ray Conf.}}
\newcommand{\xxvicrc}{{\it 25th Internat. Cosmic Ray Conf.}}
\newcommand{\mic}{$\mu$m}

\begin{document}
{\tt referee}
   \thesaurus{06       
              (03.11.1; 
               16.06.1; 
               19.06.1; 
               19.37.1; 
               19.53.1; 
               19.63.1)}
\title{Absorption of Very High Energy Gamma-Rays
by Intergalactic Infared Radiation: A New Determination}

\author{F.W. Stecker\inst{1} \and O.C. de Jager\inst{2}
          }

   \offprints{F.W. Stecker}

   \institute{Laboratory for High Energy Astrophysics, \\
	      NASA Goddard Space Flight Center, \\	
 	     Greenbelt, MD 20771, USA \and  		
             Space Research Unit, Dept. of Physics, 
              Potchefstroom University for CHE, \\
              Potchefstroom 2520, South Africa
             }  


   \maketitle

   \begin{abstract}
We present a new calculation of the intergalactic \gray\ pair-production 
absorption coefficient as a function of both energy and redshift. 
In reexamining this problem, we make use of a new {\it empirially} based 
calculation (as opposed to previous {\it model} calculations) of the 
intergalactic infrared radiation field (IIRF).
We find smaller opacities than those given previously (Stecker \& De Jager 
1997). We apply our results to the new observations of
the flaring \gray\ spectra of Mrk 421 and Mrk 501, both at a redshift of
$\sim$0.03. Our new calculations indicate
that there should be no significant curvature in the spectra of these
sources for energies below 10 TeV, 
as indicated by recent observations. However, 
the intrinsic spectra of these sources should be harder
by amounts of $\sim$ 0.25 to 0.45 in the spectral index (in the 1 - 10 TeV
range), with an intergalactic absorption cutoff above $\sim 20$ TeV.

   \end{abstract}

\section{Introduction}
We have previously pointed out (Stecker, De Jager \& Salamon 1992 
(hereafter,
SDS92)) that very high energy \gray\ beams from blazars can be used to 
measure the intergalactic infrared radiation field, since 
pair-production interactions of \grays\ with intergalactic IR photons 
will attenuate the high-energy ends of blazar 
spectra. Determining the intergalactic IR field, in turn, allows us to 
model the evolution of the galaxies which produce it. 
As energy thresholds are lowered 
in both existing and planned ground-based
air Cherenkov light detectors (Cresti 1996), cutoffs in the \gray\ spectra of 
more distant blazars are expected, owing to extinction by the IIRF. These
can be used to explore the redshift dependence of the IIRF (Stecker \& Salamon
1997; Salamon \& Stecker 1998). 
Furthermore, by using blazars for a determination 
of attenuation as a function of
redshift,  combined with a direct observation of the IR background from
the {\it DIRBE}  detector on {\it COBE}, one can, in principle, 
measure of the Hubble
constant $H_{0}$ at truly cosmological distances (Salamon, Stecker \&
De Jager 1994). 

There are now over 50 grazars which have been detected by the {\it EGRET} team
(Thompson, \etal\ 1996). These sources, optically violent variable quasars
and BL Lac objects, have been detected out to a redshift greater that 2.
Of all of the blazars detected by {\it EGRET}, only the low-redshift 
BL Lac, Mrk 421, has been seen by
the Whipple telescope. The fact that the Whipple team did not detect the
much brighter {\it EGRET} source, 3C279, at TeV energies (Vacanti, \etal\ 1990,
Kerrick, \etal\ 1993) is consistent with the predictions of a
cutoff for a source at its much higher redshift of 0.54 (see SDS92).
So too is the recent observation of two other very close BL Lacs ($z < 0.05$),
{\it viz.}, Mrk 501 (Quinn, \etal\ 1996) and 1ES2344+514 
(Catanese, \etal\ 1997) 
which were too faint at GeV energies to be seen by {\it EGRET}.

In this paper, we calculate the absorption coefficient of intergalactic
space using a new, empirically based calculation
of the spectral energy distribution (SED) of intergalactic low energy 
photons (Malkan \& Stecker 1998; hereafter MS98) 
obtained by integrating luminosity dependent infrared spectra of galaxies
over their luminosity and redshift distributions.
After giving our results on the \gray\ optical depth as a function of energy 
and redshift out to a redshift of 0.3, we apply our calculations by
comparing our results with recent spectral data on Mrk 421 as reported by 
McEnery, \etal\ (1997) and spectral data on Mrk 501 given by Aharonian,
\etal\ (1997). The results presented here supercede those of our 
previous calculations (Stecker \& De Jager 1997), which were based 
more on theoretical models (see discussion in MS98).
We consider the results presented here to be considerably more reliable
than any presented previously.

\section{The Opacity of Intergalactic Space to the IIRF}
The formulae relevant to absorption calculations involving pair-production 
are given and discussed in SDS92.
For $\gamma$-rays in the TeV energy range, the pair-production cross section 
is maximized
when the soft photon energy is in the infrared range: 
$$\lambda (E_{\gamma}) \simeq \lambda_{e}{E_{\gamma}\over{2m_{e}c^{2}}} =
2.4E_{\gamma,TeV} \; \; \mu m \eqno{(1)}$$ where $\lambda_{e} = h/(m_{e}c)$ 
is the Compton wavelength of the electron.
For a 1 TeV $\gamma$-ray, this corresponds to a soft photon having a
wavelength  near the K-band (2.2\mic). (Pair-production interactions actually
take place with photons over a range of wavelengths around the optimal value as
determined by the energy dependence of the cross section.) 
If the emission spectrum of
an extragalactic source extends beyond 20 TeV, then the extragalactic
infrared field should cut off the {\it observed} spectrum between $\sim
20$ GeV and $\sim 20$ TeV, depending on the redshift of the source (Stecker \&
Salamon 1997; Salamon \& Stecker 1998, and this paper).  

In our calculations, we
make the reasonable simplifying assumption 
that the IIRF is basically in
place at redshifts $<$ 0.3, having been produced primarily at higher
redshifts (Madau 1996; Stecker \& Salamon 1997; Salamon \& Stecker 1998
and references therein). 
We therefore limited our calculations to $z<0.3$. For a treatment of 
intergalactic absorption at higher redshifts by optical and UV photons
using recent data on galaxy evolution at moderate and high redshifts,
see Stecker \& Salamon (1997) and Salamon \& Stecker (1998).

\begin{figure}[htbp]
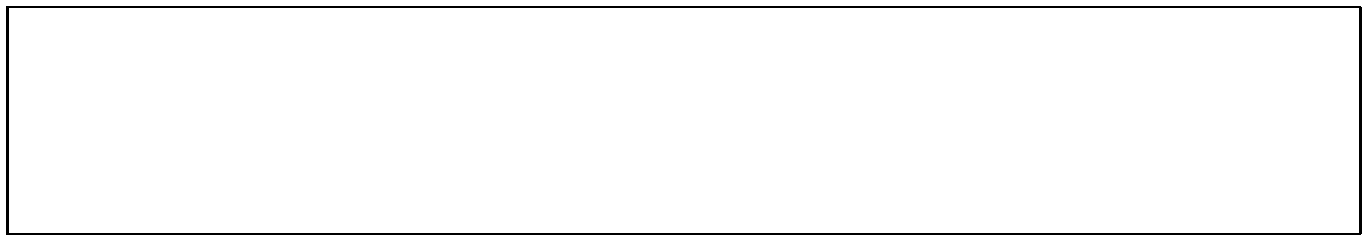

\picplace{3cm}
\caption{
The spectral energy distribution (SED) of the extragalctic 
IR radiation calculated
by Malkan \& Stecker (1997) with the 2.7 K cosmic background radiation 
spectrum added. The solid line (lower IIRF curve) and the dashed line 
(higher IIRF curve) correspond to the middle and upper curves calculated by 
Malkan \& Stecker (1997) with redshift-evolution assumptions as 
described in the text.}

\end{figure}

We assume for the IIRF, two of the SEDs given in MS98 (shown in Figure 1); 
the lower curve in Figure 1 assumes evolution 
out to $z=1$, whereas the upper curve assumes evolution out to $z=2$.
Evolution in stellar emissivity is expected to level off or decrease
at redshifts greater than $\sim 1.5$ (Fall, Charlot \& Pei 1996, Madau 1996), 
so that the two curves in Fig. 1 may be considered to be lower and upper 
limits, bounding the expected IR flux (MS98).
Using these two SEDs for the IIRF, we
have obtained parametric expressions for $\tau(E,z)$ for $z<0.3$, taking a 
Hubble constant of $H_o=65$ km s$^{-1}$Mpc$^{-1}$ (Gratton, \etal\ 1997). 
The double-peaked form of the SED of the IIRF requires
a 3rd order polynomial to reproduce parametrically. It is of the form
$$log_{10}[\tau(E_{\rm TeV},z)]\simeq\sum_{i=0}^3a_i(z)(\log_{10}E_{\rm TeV})^i
\;\;{\rm for}\;\;1.0<E_{\rm TeV}<50, \eqno(2)$$
where the z-dependent coefficients are given by
$$a_i(z)=\sum_{j=0}^2a_{ij}(\log_{10}{z})^{j}. \eqno(3)$$

Table 1 gives the numerical values for 
$a_{ij}$, with $i=0,1,2,3$, and $j=0,1,2$. The numbers before the
brackets are obtained using the lower IIRF SED shown in Figure 1; The
numbers in the brackets are obtained using the higher IIRF SED.
Because we are using real IRAS data to give more accurate estimates of the 
IIRF, we do not give values of $\tau$ for E $<$ 1 TeV
and for larger redshifts, which would involve interactions with 
starlight photons of wavelengths $\lambda \le 1\mu$m. Equation (2) approximates
$\tau(E,z)$ correctly within 10\% for all values of $z$ and $E$ considered.

\vspace{1em}

\begin{tabular}{|c|r|r|r|r|}
\multicolumn{5}{c}{\bf Table 1: Polynomial coefficients $a_{ij}$}\\
\hline
$j$ & $a_{0j}$ & $a_{1j}$ & $a_{2j}$ & $a_{3j}$ \\ \hline
0&1.11(1.46) &-0.26(~0.10) &1.17(0.42) &-0.24(~0.07)\\
1&1.15(1.46) &-1.24(-1.03) &2.28(1.66) &-0.88(-0.56)\\
2&0.00(0.15) &-0.41(-0.35) &0.78(0.58) &-0.31(-0.20)\\ \hline
\end{tabular}

\begin{figure}[htbp]
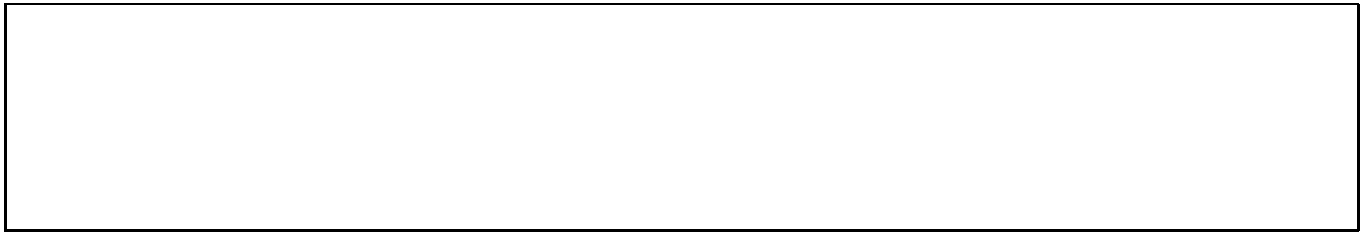

\picplace{3cm}
\caption{
Optical depth versus energy for \grays\ originating at various redshifts
obtained using the SEDs corresponding to the lower IIRF (solid lines) and 
higher IIRF (dashed lines) levels shown in Fig. 1.}
\end{figure}

\begin{figure}[htbp]
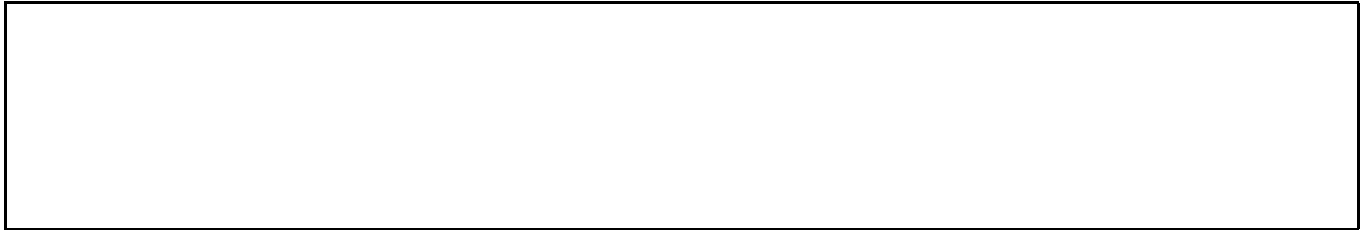

\picplace{3cm}
\caption{
The observed spectra of Mrk 421 from McEnery, \etal\ (1997) (open triangles)
and Mrk 501 from Aharonian, \etal\ (1997) (solid circles - spectrum divided 
by 10). Best-fit absorbed spectra (of the form $KE^{-\Gamma}\exp(-\tau(E,z=0.03))$)
and unabsorbed spectra ($KE^{-\Gamma}$) for both sources
are shown for $\tau$ corresponding to
the lower IIRF SED (solid lines; $\Gamma=2.36$ and 2.2 for Mrk 421 and 
Mrk 501 respectively) and higher IIRF SED (dashed lines;
$\Gamma=2.2$ and 2.03 for Mrk 421 and Mrk 501 respectively).}
\end{figure}

\vspace{1em}
Figure 2 shows the results of our calculations of the optical depth for 
various energies and redshifts up to 0.3.
Figure 3 shows observed spectra 
for Mrk 421 (McEnery, \etal\ 1997) and Mrk 501 (Aharonian \etal\ 1997) 
in the flaring phase, compared with best-fit spectra of the form
$KE^{-\Gamma}\exp(-\tau(E,z=0.03))$, with $\tau(E,z)$ given by the two 
appropriate curves shown in Figure 2. Because $\tau < 1$ for $E<10$, 
TeV, there is no obvious curvature in the
{\it differential} spectra below this energy; rather, we obtain a
slight steepening in the power-law spectra of the sources as a result 
of the weak absorption. This result implies that the 
{\it intrinsic} spectra of the sources should be harder by 
$\delta \Gamma \sim$ 0.25 in the 
lower IRRF case, and $\sim$ 0.45 in the higher IIRF case. 

\section{Discussion and Conclusions}
We have calculated the absorption coefficient of intergalactic space
from pair-production interactions with low energy photons of the IIRF,
both as a function of energy and redshift, using new, more reliable estimates 
of the SED for the IIRF which were obtained by MS98.
Our results
predict less absorption than we obtained previously (Stecker \& De Jager
1997), because the MS98 IIRF is lower than our previous estimate.
One reason for this difference is that our previous IIRF spectrum was 
normalized partly to
reflect the estimate of Gregorich, {\it et al.} (1995) of the IIRF at 60 \mic.
According to Bertin, Dennefeld \& Moshir (1997), that estimate may have been 
based on the inclusion of false detections in their analysis of the IRAS data.
For absorption calculations, it is important to use the most 
reliable estimate of the IIRF avaliable, since
the absorption effect depends exponentially on the magnitude of the IIRF.
While we do not claim that our new results for the absorption coeeficient as 
a function of energy differ dramatically from those obtained
previously (MacMinn \& Primack 1996; Stecker \& De Jager 1997), we {\it do} 
claim that they are more reliable because they are
based on the empirically derived IIRF given by MS98, whereas all previous
calculations of TeV $\gamma$-ray absorption were based on theoretical modeling
of the IIRF. 
The MS98 calculation was based on data from nearly 3000
IRAS galaxies. These data included (1) the luminosity dependent infrared SEDs
of galaxies, (2) the 60$\mu$m luminosity function of galaxies and, (3) 
the redshift distribution of galaxies. 

We have applied our absorption calculations to recent flaring spectra of the 
nearby BL Lac objects Mrk 421 and Mrk 501. 
The spectral calculations given here are in good agreement with the
recent observations that indicate no significant 
curvature in the spectra of Mrk 421 and Mrk 501. The observations are also 
consistent with our calculated steepening of 0.25 to
0.45 in the spectral index of these sources in the 1-10 TeV range.
Our new calculations predict a significant intergalactic absorption effect
which should cut off the spectra of Mrk 421 and Mrk 501 at energies greater
than $\sim$20 TeV. Observations of these objects at large zenith angles, 
which give large effective threshold energies, may thus demonstrate 
the effect of intergalactic absorption.

Our new calculations confirm the conclusion in SDS92 that
TeV spectra of sources at redshifts higher than 0.1 should suffer 
significant absorption. The recent detection of another 
XBL at a redshift below 0.1, {\it viz.}, 1ES2344+514 (Catanese, \etal\ 1997),
further supports the argument that nearby XBLs may be the only significant 
TeV sources presently detectable (Stecker, De Jager \& Salamon 1996).

\begin{acknowledgements}
We wish to acknowledge R. Lamb, F. Aharonian and G. Hermann for sending us 
their \gray\ data in numerical form. We also wish to thank P. Fleury for 
helpful criticism of the manuscript.

\end{acknowledgements}

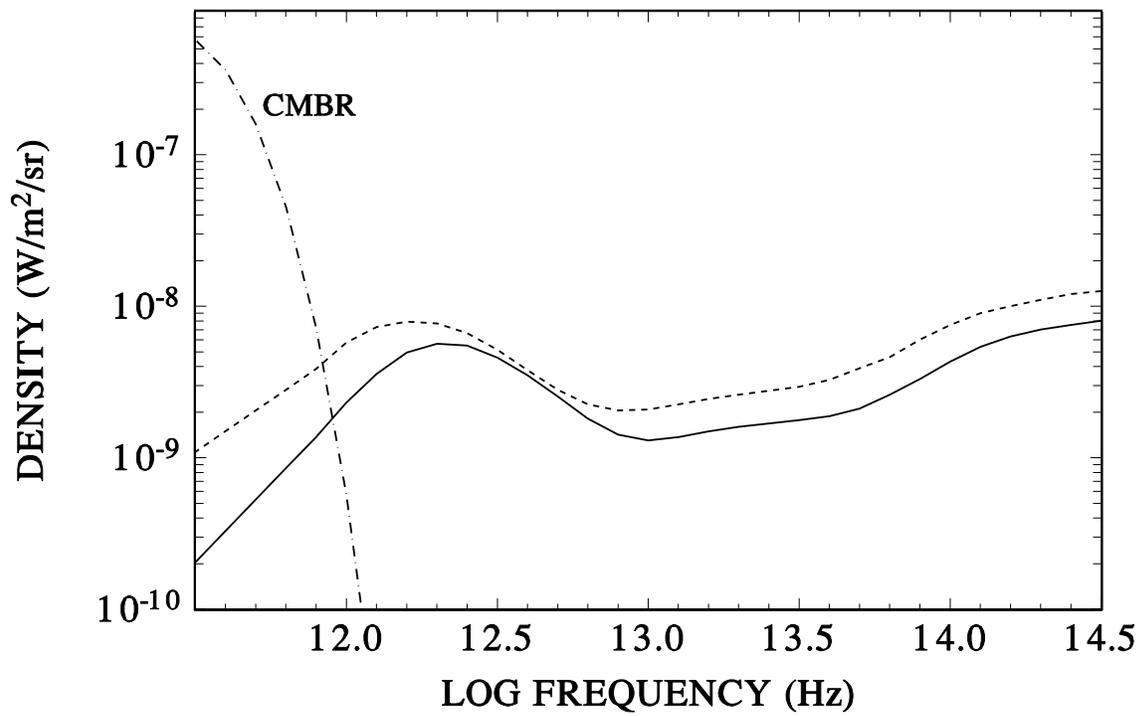
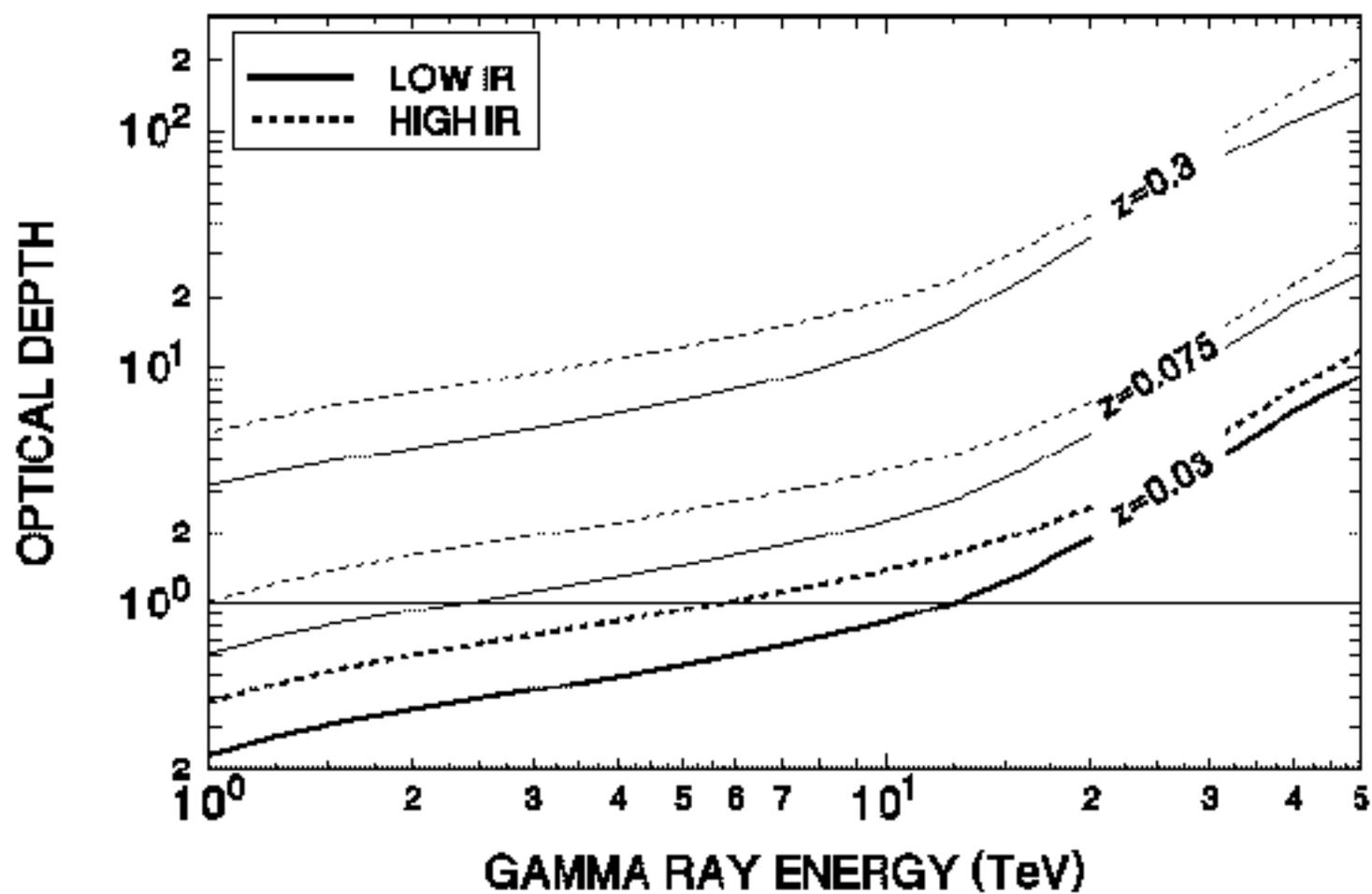
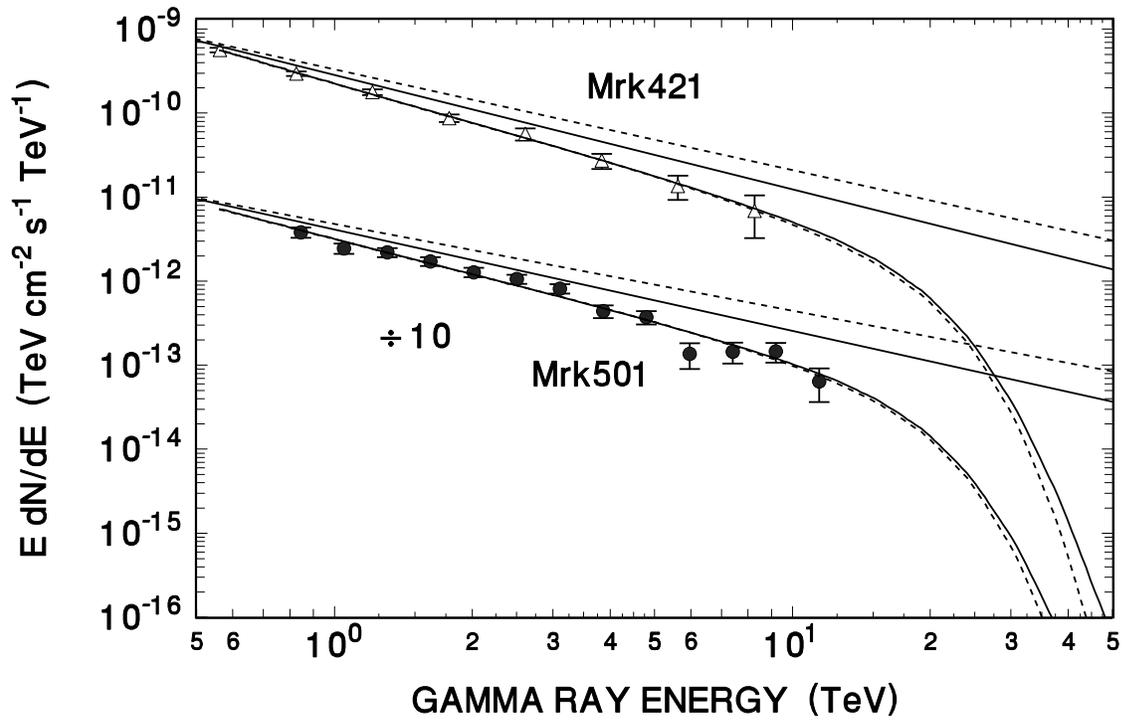
\end{document}